\documentclass{article} 
\usepackage{iclr2026_conference,times}

\usepackage{amsmath,amsfonts,bm}

\def\eqref#1{equation~\ref{#1}}

\def\1{\bm{1}}

\DeclareMathAlphabet{\mathsfit}{\encodingdefault}{\sfdefault}{m}{sl}
\SetMathAlphabet{\mathsfit}{bold}{\encodingdefault}{\sfdefault}{bx}{n}

\usepackage{hyperref}
\usepackage{url}
\usepackage{algorithm}
\usepackage{algpseudocode}
\usepackage{graphicx}

\title{A Fully Differentiable \\Framework for training proxy\\ Exchange Correlation Functionals\\ 
For Periodic systems}

\author{Rakshit Kumar Singh, Aryan Amit Barsainyan \& Bharath Ramsundar \\Deep Forest Sciences \\
\texttt{\{rakshit,aryan,rbharath\}@deepforestsci.com} \\}
\iclrfinalcopy 
\begin{document}

\maketitle

\begin{abstract}

Density Functional Theory (DFT) is widely used for first-principles simulations in chemistry and materials science, but its computational cost remains a key limitation for large systems. Motivated by recent advances in ML-based exchange–correlation (XC) functionals, this paper introduces a differentiable framework that integrates machine learning models into density functional theory (DFT) for solids and other periodic systems. The framework defines a clean API for neural network models that can act as drop in replacements for conventional exchange-correlation (XC) functionals and enables gradients to flow through the full self-consistent DFT workflow. The framework is implemented in Python using a PyTorch backend, making it fully differentiable and easy to use with standard deep learning tools. We integrate the implementation with the DeepChem library to promote the reuse of established models and to lower the barrier for experimentation. In initial benchmarks against established electronic structure packages (GPAW and PySCF), our models achieve relative errors on the order of 5–10\%.

\end{abstract}

\section{Introduction}

Density Functional Theory (DFT) has been central to modern computational chemistry and materials science, enabling first-principles simulations across diverse systems, but its high computational cost remains a major limitation, especially for large-scale systems. This has motivated the development of machine learning surrogate models that approximate DFT level accuracy at a fraction of the computational cost. Although recent advances in deep learning have significantly advanced ML based proxy models, many existing implementations remain tightly coupled to legacy FORTRAN/C based calculators or lack support for periodic boundary conditions, creating fragmentation that hinders integration with modern ML frameworks.

\cite{vidhyadhiraja2023opensourceinfrastructuredifferentiable} introduced an open source infrastructure for differentiable density functional theory (DFT) to enable optimization of neural exchange correlation (XC) functionals, through the implementation of differentiable DFT components. This infrastructure was integrated into DeepChem \cite{Ramsundar-et-al-2019}, an open source Python library for machine learning on molecular, biological, and quantum datasets, built on PyTorch and designed to provide a unified ecosystem for model development, training, and inference. Together, this integration enabled direct optimization of learned functionals using physically meaningful objectives within a modern, extensible ML framework.

Building on this foundation, we further extend DeepChem to support Kohn Sham DFT for periodic systems through a native solid-state representation (the \texttt{Sol} class), enabling end to end learning and optimization of XC functionals for crystalline materials. In this work, we present an python implementation of DFT, which is modular enough to work as a framework for building proxy ML models of Exchange Correlation functionals. It maintains flexibility and interoperability by keeping most of the computation on PyTorch backend, with a future possibility of GPU acceleration. A key design goal is to have seamless support for Periodic Boundary Conditions, enabling modeling of bulk materials and surfaces without doing approximations on a molecular system. We have integrated all the code into deepchem for easy use and long term maintainability.

\section{Related Work}

In recent years, research has increasingly shifted toward physically grounded machine learning approaches that integrate directly with the Kohn Sham formalism, rather than bypassing it. A central direction is to replace the exchange correlation (XC) functional itself with machine learning models, beginning with density functional learning for simple systems \cite{PhysRevLett.108.253002} and later incorporating physical structure through Kohn Sham based regularization \cite{PhysRevLett.126.036401}. However, many such approaches treat the self consistent field (SCF) procedure as a black box, limiting optimization and preventing gradient propagation through the Kohn Sham equations. Recent advances in automatic differentiation have addressed this limitation by enabling fully differentiable quantum chemical pipelines, from variational Hartree–Fock \cite{Tamayo_Mendoza_2018} to differentiable DFT frameworks that preserve physical structure by learning only the XC term while keeping other contributions physics-based \cite{PhysRevLett.127.126403}, establishing a principled paradigm for machine learning driven exchange correlation modeling. A recent work has further advanced this direction by unifying variational principles, automatic differentiation, and neural representations within end to end differentiable DFT frameworks, reinforcing the emerging paradigm of physics informed, fully differentiable electronic structure modeling and aligning closely with the direction pursued in the present work. \cite{PhysRevLett.133.076401}

\section{Methods}

\begin{figure*}[t]
    \centering
    \makebox[\textwidth][c]{%
        \includegraphics[width=1.2\textwidth]{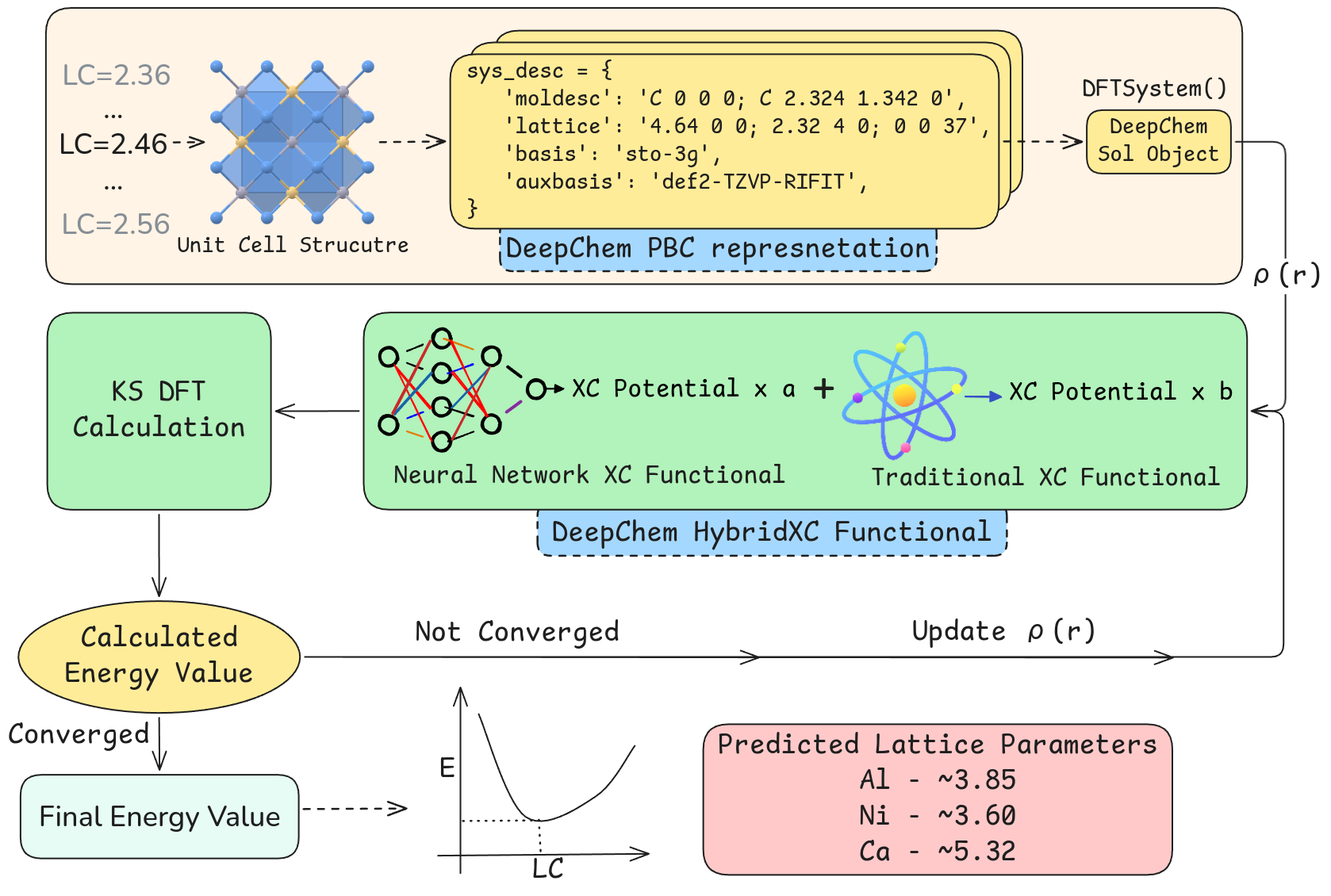}
    }
    \caption{This figure represents the Lattice constant optimization flow using a trained NN XC functional used alongside a DFT XC functional. The framework was tested using lattice constants from experimental data as initial guess of the system.}
    \label{fig:placeholder}
\end{figure*}

We employ a multi layer perceptron (MLP) coupled it with a generalized gradient approximation (GGA) functional through the \texttt{HybridXC} class, enabling a hybrid formulation. The resulting hybrid functional is trained on periodic atomic structures. This hybrid functional is then used within a periodic DFT workflow to evaluate total energies across varying lattice parameters.

\subsection{Dataset}

We focused our efforts on Face Centered Cubic systems as they are abundant in nature and provide good test candidates for our machine learning tool. The experimental reference data for this paper were obtained from Materials Project. \cite{10.1063/1.4812323} The dataset processing pipeline is detailed in appendix \ref{Data Generation}.

While \cite{vidhyadhiraja2023opensourceinfrastructuredifferentiable} represented molecular systems using basis-set information and molecular descriptors (\texttt{moldesc}), enabling a dictionary-based construction of the initial electron density, this representation is not directly transferable to periodic systems. Periodic calculations require explicit encoding of lattice geometry and auxiliary basis information associated with the unit cell. To preserve compatibility with the existing DeepChem-based framework while extending it to crystalline systems, we introduce a minimal but systematic modification by augmenting the system representation with a \texttt{lattice} keyword argument, enabling the inclusion of lattice vectors and periodic cell information without altering the core data structure or workflow design. This representation enables consistent periodic calculations required for lattice constant optimization.

\subsection{Lattice Parameter Optimization Workflow}

Using this representation and hybrid functional, we perform lattice parameter optimization as follows. We begin by selecting an initial estimate of the lattice constant. For validation and testing purposes, this initial value is taken from available experimental data. However, the starting structure can also be obtained from a prior force-relaxed configuration generated through molecular dynamics simulations or other computationally inexpensive relaxation methods. To locate the minimum energy region with respect to the lattice constant, we define a search interval around the initial estimate. Density Functional Theory (DFT) calculations are then performed at each sampled lattice constant using our hybrid DFT workflow. The total energy corresponding to each lattice parameter is recorded, allowing us to construct an energy versus lattice constant curve. The equilibrium lattice constant is determined from the minimum of this energy curve. If the minimum does not lie within the initially chosen bandwidth, the search interval is extended in the direction where the energy continues to decrease, and the scanning procedure is repeated until the minimum is captured.

\section{Results and Observations}

\begin{figure*}[t]
    \centering
    \makebox[\textwidth][c]{
    \includegraphics[width=0.9\textwidth]{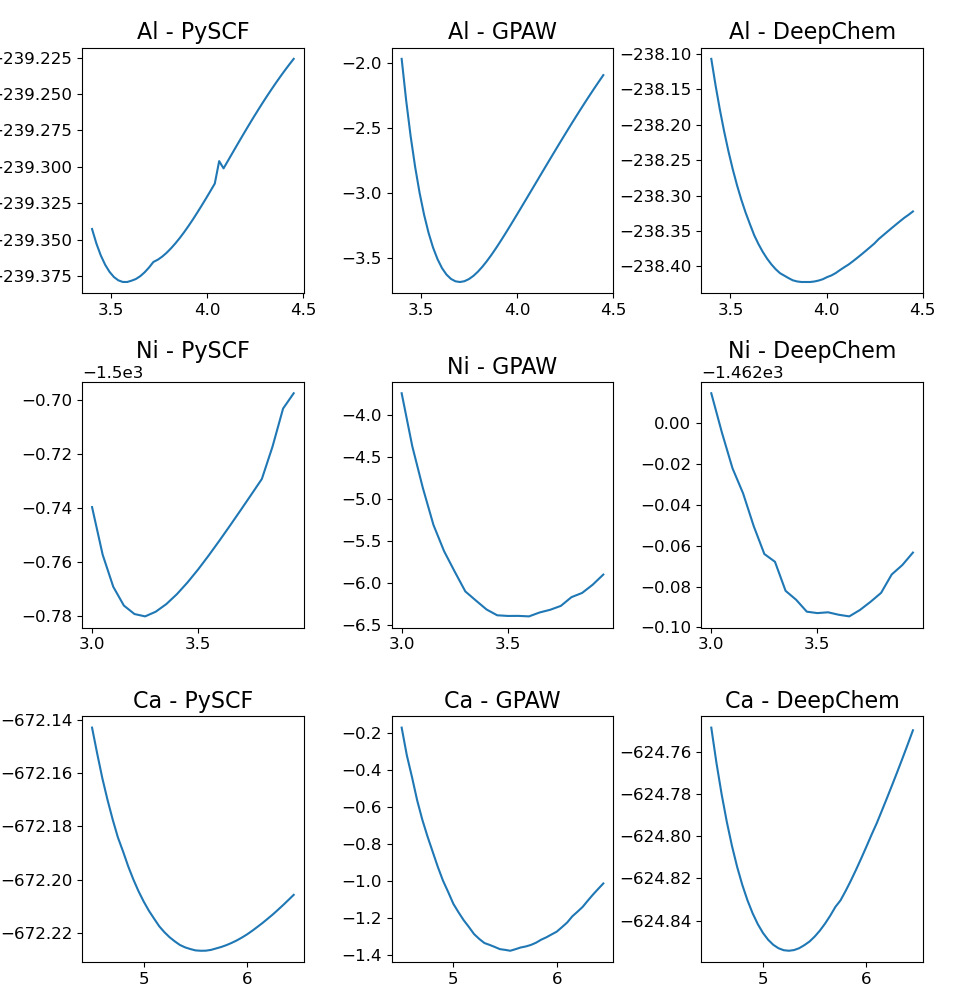}
    }
    \caption{This figure presents the lattice constant scans for Al, Ni, and Ca computed using PySCF, GPAW, and DeepChem. The lattice parameters were varied from $3.40 \AA$ to $4.45 \AA$ for Al, $3.00 \AA$ to $4.00 \AA$ for Ni, and $4.50 \AA$ to $6.50 \AA$ for Ca, with a step size of $0.05 \AA$.}
    \label{RESULTS}
\end{figure*}

The proposed DFT+NN framework was trained on a subset of the Materials Project dataset (see Appendix ~\ref{Reference Training Data}). The trained model was subsequently evaluated on aluminium (Al), nickel (Ni), and calcium (Ca), with the corresponding results shown in Fig.~\ref{RESULTS}. For these systems, the predicted lattice constants show a relative error of approximately 5–10\% compared with reference DFT calculations performed using established packages such as GPAW and PySCF.

However, the framework does not yield physically meaningful predictions for graphene (Appendix.~\ref{Graphene}), highlighting current limitations of the model when applied to certain classes of materials.

\section{Conclusion}

Training DFT-ML models using a hybrid combination of a traditional exchange correlation (XC) functional and a neural network provides the model with a strong physical starting point rather than forcing it to learn electronic structure behavior from scratch. The classical XC functional already captures well established physical principles and correct qualitative trends, allowing the neural network to focus on learning systematic deficiencies instead of reproducing known physics. This setup makes training more stable, reduces the amount of data required, and helps prevent nonphysical behavior during self-consistent iterations. By grounding the learning process in a reliable physical baseline, the model remains robust and interpretable while still gaining flexibility from machine learning.

\section{Future Work}

A natural extension of the present differentiable DFT framework is the adoption of a plane wave basis. Plane waves provide a systematic representation of periodic systems through a complete, orthogonal basis controlled by a single energy cutoff. Unlike localized atomic orbitals, they avoid basis set superposition errors due to their uniform, position independent nature. Their translational invariance makes them particularly robust for metallic and low dimensional systems with delocalized electronic states. Consequently, plane wave DFT offers a stable and scalable foundation for learning transferable exchange correlation models in solids within a fully differentiable framework. \cite{RevModPhys.64.1045}

\bibliography{iclr2026_conference}
\bibliographystyle{iclr2026_conference}

\clearpage
\appendix
\section{Appendix}

\subsection{Data Generation}
\label{Data Generation}

Training data are generated from self consistent Kohn Sham DFT calculations using periodic atomic structures as input, defined by atomic positions $R_{N \times 3}$ and unit cell lattice vectors $L_{3 \times 3}$. For each structure, a self-consistent field (SCF) calculation is performed using a well established exchange correlation functional (LDA or GGA) until convergence, yielding the total ground-state energy. The total energy is expressed as
\begin{equation}
E_{\text{tot}}^{\text{DFT}} =
E_{e\text{-}core} +
E_{e\text{-}e} +
E_{n\text{-}n} +
E_{xc}^{\text{DFT}}(\rho),
\end{equation}
where $E_{e\text{-}core}$ denotes electron-core interactions, $E_{e\text{-}e}$ electron-electron interactions, $E_{n\text{-}n}$ nuclear-nuclear repulsion, and $E_{xc}^{\text{DFT}}(\rho)$ the exchange-correlation energy functional of the electron density $\rho$. After convergence of the SCF cycle, the total energy $E_{\text{tot}}$ is extracted and used as the reference target for model training. Each data sample is therefore represented as
\begin{equation}
\mathcal{D} = \{ R, L, E_{\text{tot}} \}.
\end{equation}

All reference DFT calculations are performed using DeepChem's DFT implementation to ensure strict numerical and methodological consistency. This is critical because DFT results depend not only on atomic structure, but also on implementation specific choices such as basis sets, numerical grids, and SCF convergence schemes, which can introduce systematic discrepancies across different codes.

\subsection{Training Objective}

The neural network is trained to reproduce reference DFT total energies by minimizing the loss function:

\begin{equation}
\mathcal{L}(\theta)
=
\frac{1}{N}
\sum_{i=1}^{N}
\frac{(E_{tot,i}^{DFT}
-
E_{tot,i}^{Hybrid}
)^2}{E_{tot,i}^{DFT}}
\end{equation}

where:

\begin{equation}
E_{tot}^{NN}
=
E_{e-core}
+
E_{e-e}
+
E_{n-n}
+
E_{xc}^{NN}(\rho)
\end{equation}

During training, the Hartree, electron–core, and nuclear–nuclear energy contributions are computed using physics, while only the XC term is learned by the neural network.

\subsection{Data Representation}

\[
\begin{array}{l}
\texttt{sys\_desc = \{} \\
\texttt{\ \ 'moldesc': 'C 0 0 0; C 2.324 1.342 0',} \\
\texttt{\ \ 'lattice': '4.64 0 0; 2.32 4 0; 0 0 37'} \\
\texttt{\ \ 'basis': 'sto-3g',} \\
\texttt{\ \ 'auxbasis': 'def2-sv(p)',} \\
\texttt{\}}
\end{array}
\]

\subsection{K-Point Sampling}

When doing DFT on periodic systems, we described electronic states by Bloch functions due to the translational invariance of the Kohn Sham Hamiltonian, and physical observables are formally defined as integrals over the first Brillouin zone. In practical calculations, these integrals are evaluated numerically using finite k-point sampling in reciprocal space. Accordingly, our periodic DFT calculations employ standard Brillouin zone k-point meshes during the self-consistent field procedure, with energies, densities, and related quantities computed as weighted sums over sampled k-points.

\subsection{Limitations Observed in Graphene Calculations}
\label{Graphene}

We observed an opposite trend in the total energy relaxation curve when compared to reference results obtained using GPAW. Notably, similar behavior was also present in calculations performed with our tool without the inclusion of the neural network component, suggesting that the issue does not originate from the learned exchange correlation model but rather from the underlying DFT implementation. This indicates that the base framework may require further tuning to reliably handle more complex periodic systems. While the use of more advanced exchange correlation functionals and larger basis sets improved numerical stability, they did not lead to a corresponding improvement in accuracy. Based on these observations, we suspect that transitioning to a plane wave based formulation, which is better suited for periodic systems, could resolve many of these issues.

\subsection{Reference Training Data}
\label{Reference Training Data}

The reference dataset was generated using density functional theory calculations with the GGA exchange--correlation functional, specifically the PBE formulation (gga\_x\_pbe + gga\_c\_pbe), in combination with the STO-3G Gaussian basis set. The resulting energies serve as ground truth training data for model development and validation, and as benchmarking references for evaluating the accuracy of the proposed periodic DFT framework.

\begin{table}[h]
\centering

\label{tab:total_energies}
\begin{tabular}{llc}
\hline
\textbf{System} & \textbf{Lattice Constant} \\
\hline
Cu & 3.615 \\
Ca & 5.58 \\
Ag & 4.09 \\
Pd & 3.89 \\
Rh & 3.80 \\
Al & 4.05 \\
Ni & 3.52 \\
MgO  & 2.661 \\
CaF$_2$ & 3.900 \\
\hline
\end{tabular}
\end{table}

\subsection{Additional Note}

We observed some inconsistencies in the plot generated by pyscf at 3.70 $\AA$, 3.75$\AA$, 3.95$\AA$ and 4.00 $\AA$ lattice constants. This does not affect the smooth minima we obtained in the start of the plot, hence we do not include these datapoints in the fig. \ref{RESULTS}. The main reason we suspect for this behavior is that we were using strict and small basis sets and xc functionals on all the tests.

\begin{figure}[H]
    \centering
    \includegraphics[width=0.35\columnwidth]{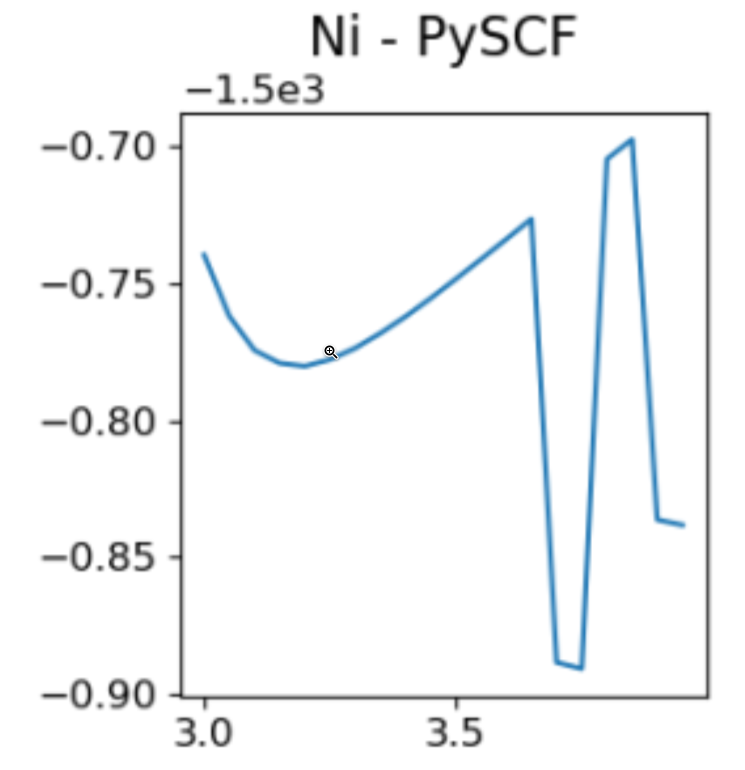}
    \caption{Original PySCF graph}
    \label{RESULTS}
\end{figure}
\end{document}